\documentstyle[fleqn,twoside,gc]{article}

\heads{J.A. de Freitas Pacheco and S. Peirani}
      {Dark Matter: indirect detection }

\begin{document}
\twocolumn[
%\Arthead{6}{2005}{X (XX)}{X}{XX}
%\Arthead{?}{2005}{? (?)}{?}{?}
\Arthead{?}{2005}{? (?)}{0}{0}

\Title{Indirect Search for Dark Matter}

\Author{J.A. de Freitas Pacheco            %%
              and S. Peirani}   %%   If there is
 {Observatoire de la C\^ote d'Azur, BP4229, F06304, Nice Cedex 4, France}              %%   a single address

\Abstract
    {Possible dark matter candidates are reviewed as well as indirect
search methods based on annihilation or decay channels of these particles. 
Neutralino is presently the best particle candidate and its annihilation produces high energy neutrinos, antiprotons, 
positrons and $\gamma$-rays. To date, only
upper limits on neutrino fluxes from the center of the Earth or the Sun, 
were established by different experiments. Antiprotons detected by the BESS
collaboration, if issued from the follow up hadronization of the annihilation process, exclude
neutralino masses higher than 100 GeV. The EGRET $\gamma$-ray residual emission seen
at high galactic latitudes above 1 GeV could be explained by neutralino annihilations
if: i) the dark matter profile is ``cored" and ii) the neutralino mass is $\leq$ 50 GeV.
Sterile neutrinos in the keV mass range are a possible candidate to constitute warm dark matter. These 
particles
may provide an adequate free streaming mass able to solve {\it some} 
difficulties present in the cold dark matter scenario at small scales and could also explain the
natal kick of pulsars.  
MeV particles, dubbed {\it light} dark matter, proposed to explain
the extended 511 keV line emission from the galactic center will also be discussed.}

\RAbstract               %% (necessary for Russian-speaking authors *)
    {Title in Russian}
    {Author(s) in Russian}
    {Text of abstract in Russian}

%%% *  For foreign authors, a Russian translation of the Abstract
%%%    is provided by the editors.

]  %%%%%%%%%%%%%%%%%%%%%%%%%%%%%  End of temporary one-column mode
\email 1 {pacheco@obs-nice.fr}
\email 2 {peirani@obs-nice.fr}

\section{Introduction}

Baryons represent only a minor percentage ($\sim$ 4\%) in the matter-energy budget of the universe, the 
remaining and major part being probably under the form
of some kind of ``exotic" matter. Data on angular power spectrum of temperature fluctuations
of the cosmic microwave background radiation derived from WMAP and on the luminosity-distance 
of type Ia supernovae [1,2,3], indicate that the ``exotic" matter has in fact
two components: one, which acts as a ``repulsive" force, labeled {\it dark energy}
and another, which is responsible for gravitational forces at large scales, dubbed
{\it dark matter}. The former corresponds to about 70\% whereas the latter corresponds
to about 26\% of the total matter-energy content of the universe. 
The dark energy component, sometimes identified as the ``cosmological constant" ($\Lambda$) term,
first introduced by Einstein, is
responsible for the observed acceleration of the expansion of the universe [2,3].
Because of conceptual problems associated with the so-called ``$\Lambda$"-term, different
alternatives have been explored in the literature. The most popular, christened ``quintessence",
uses a scalar field $\phi$ with a suitable potential $V(\phi)$ so as to make the
vacuum energy density vary with time. However,
the possible nature of the dark energy will not be discussed in this paper, whose main purpose 
is to find answers to the question: {\it what is dark matter made of} ?

Among particles issued from the Standard Model, the only particle which has an important
relic density is the neutrino. However, recent observational constraints obtained from
WMAP data [1] imply that their total mass density should satisfy $\Omega_{\nu}h^2 < 0.0076$,
considerably less than the amount of gravitational mass present in the universe
($\Omega_mh^2 = 0.13\pm 0.02$). Moreover, neutrinos are relativistic at the freeze-out
and due to their relativistic streaming, small-scale structures are erased, difficulting
the formation of galaxies and ruling out neutrinos as an acceptable dark matter candidate.

Axions and massive Higgs-like bosons have also been proposed in the 
past as dark matter candidates.
Presently, we do not know either the relic abundance or the interaction type besides
gravitation to which these particles are subjected. Very massive boson fields may have
played an important r\^ole in the formation of the present observed large structure of
the universe, since they may experience gravitational instability [4]. Moreover, boson
condensates could have been ``seeds" of primordial black holes [5], which may grow by
accreting dark and baryonic matter and are probably present today in the center of most of
galaxies. One of the difficulties to form these bosonic configurations is that if there
is no efficient cooling mechanism to get rid of the excess kinetic energy, the gravitational
collapse leads to a diffuse virialized cloud, but not a compact object. This outstanding 
problem was considered in reference [6], where the authors showed that, in fact, there is 
a dissipationless cooling
mechanism, similar to the violent relaxation of collisionless stellar systems, which leads
to the formation of compact bosonic configurations. Limits on the density of these objects
in the galactic halo were discussed in [5].

Proposed extensions of the Standard Model or Supersymmetric (SUSY) theories lead naturally to
a series of candidates, which may be point-like or not. In the former case examples
are sneutrinos, axinos, gravitinos, photinos, neutralinos, while in the latter, Q-balls
are one interesting possibility [7,8], since their self-interaction cross section may be
of the order of 20 mb or larger. These values are required for self-interacting dark
matter halo models, in order to remove the central density cusp predicted by simulations, but
not seen in the rotation curve of luminous galaxies [9]. Superheavy particles 
dubbed ``cryptons", with
masses around 10$^{14-15}$ GeV, which could have been produced non-thermally in the
very early universe, have also been proposed as a possible dark matter candidate [10]. If the
decay timescale $\tau_X$ of ``cryptons" is in the range 0.066 $\leq$ H$_0\tau_X \leq$ 1.0,
then estimates of the relic density of these particles can be made. The reasoning is the
following: high energy neutrinos can be produced by the decay of ``cryptons". Non-zero
mass very energetic neutrinos may annihilate interacting with cosmic background antineutrinos,
producing Z$^0$ gauge bosons at the resonant energy $E_r = M_Z^2/2m_{\nu}$. If the
neutrino mass is $m_{\nu} \sim$ 0.07 eV, the resonant energy is $\sim 6 \times 10^{13}$ GeV.
The Z$^0$ decay produces about 30\% of very high energy protons and 70\% of $\gamma$'s.
Since the flux of UHE protons are constrained by observations, the density of cryptons is
restricted to the
range $6 \times 10^{-10} < \Omega_X < 1.6 \times 10^{-6}$ [5], several
orders of magnitude less than the value derived from WMAP data [1].

Presently, the most plausible SUSY dark matter candidate is the neutralino ($\chi$), which is
the lightest supersymmetric particle. The neutralino is stable and hence is a candidate
relic from the Big Bang, if R-parity quantum number, introduced to avoid a too rapid
proton decay, is conserved as is the case in the Minimal Supersymmetric Extension of
the Standard Model (MSSM). The neutralino is an electrically neutral Majorana fermion
whose mass $m_{\chi}$ can range from a few GeV to few hundreds of TeV. A lower limit
of about 30 GeV has been set by the LEP accelerator [11], while an upper limit of
340 TeV is favored theoretically to preserve unitarity [12].

\section{Neutralino detection}

\subsection{Direct methods}

Direct detection of dark matter particles is based on the possibility of measuring the
recoil energy (few up to few tens of keV) of a nucleon after an elastic collision with
a putative WIMP. Since the interaction cross section is quite small ($< 10^{-6}$ pb),
large detector masses are required in order to obtain a significant event rate. The
expected low event rate demands a very low radioactive and cosmic ray background, which
is one of the major difficulties of a direct search for dark matter particles
(see reference [13] for a recent review on direct experiments). Direct
detection experiments also use the annual modulation of the signal due to the orbital
motion of the Earth around the Sun as a signature. It should be emphasized that the search strategy
and data analysis depend on the {\it assumed} spatial distribution of dark matter
and its dynamics in the galactic halo, which are not well understood yet. For instance,
it is not established if dark matter halos are presently relaxed structures or not.
Whether dark matter is homogeneously distributed with isotropic velocity distribution or
whether there are local inhomogeneities such as local streams, e.g., like that
manifested through the tidal arms of the Sagittarius dwarf, is not entirely clear.
Moreover, dark halos are generally not at rest and have considerable
angular momentum [14], whose vector direction is generally not coincident with that
of the present spin axes of baryonic disks [15]. All of these are just a few of many
uncertainties about properties of dark halos which overshadow the interpretation of
direct experiments.

\subsection{Indirect dark matter searches}

Indirect methods search for products of self-annihilation of neutralinos such as energetic
leptons, hadrons and particles emerging in the follow up hadronization and fragmentation
processes, according to the channels:

\begin{equation}
\chi\bar\chi \rightarrow l\bar l, q\bar q, W^+W^-, Z^0Z^0, H^0H^0, Z^0H^0, W^{\pm}H^{\mp}
\end{equation}

High energy neutrinos are produced either in quark jets ($b\bar b$
interactions) or in the decay of $\tau$ leptons and gauge bosons. Neutrinos
produced in the former process are less energetic than those produced in the latter.
Neutralinos can be decelerated by scattering off nuclei and then accumulating at the center
of the Earth and/or at the center of the Sun (or inside any other gravitational
potential well), thus increasing the annihilation rate.

Searches for neutrinos resulting from the above processes in the center of the Earth
have been performed by different experiments as MACRO [16], Baksan [17],
Super-Kamiokande [18] and AMANDA [19,20]. So far, these experiments have only
managed to set upper limits on neutrino fluxes coming from the center of the
Earth or from the Sun. However, many uncertainties still exist in estimates
of the capture rate of WIMPs by the Earth. New detailed numerical simulations of the 
diffusion process suffered by WIMps inside the solar system indicate
that the velocity distribution is significantly suppressed below 70 km/s
[21] (and references therein). As a consequence, the capture and the
annihilation rates are substantially
reduced if the WIMP mass is higher than $\sim$ 100 GeV. This suppression will make
the detection of neutrinos resulting from the annihilation of neutralinos in the
center of the Earth much harder when compared with previous estimates [21].

Besides high energy neutrinos, antiprotons [22,23] and positrons [24,25] are  
produced in the annihilation process too. Antiprotons are the consequence of
the hadronization of quarks and gluons whereas positrons are mainly the
result of the decay of charged gauge bosons.

Antiprotons (and positrons) are also expected to be generated by interactions
of cosmic rays with interstellar matter. However, the energy spectrum of
secondary antiprotons falls steeply for energies less than a few GeV, which
could favor the distinction between production by cosmic ray interactions and neutralino annihilation.
Antiprotons with energies in the range 0.18-1.4 GeV were detected by the balloon 
borne experiment BESS [26]. Uncertainties on the parameters characterizing our diffusive halo
(scale of the confinement region, dependence of the diffusion coefficient on energy, etc.) 
difficult analyses of such data. In spite of these unsolved problems, the present data
seem to exclude neutralino masses higher than 100 GeV [23]. 
Concerning cosmic positrons, data obtained by the High-Energy Antimatter Telescope (HEAT)[27]
suggest a slight flux excess above 5 GeV. It was shown that such an excess cannot be 
explained by annihilation of dark matter 
particles, unless a substantial number of substructures are present in the galactic
halo at a rather unlikely amount [28].

Energetic $\gamma$-rays are also produced during the neutralino annihilation process. Since
this is one of the most interesting possibilities for indirect detection of supersymmetric
matter, we will analyze this aspect in some more detail in the next section.
  
\section{$\gamma$-rays from dark halos}

The decay of neutral pions formed in the hadronization process is the dominant source of
continuum $\gamma$-rays. Besides the continuum emission, two annihilation channels may
produce $\gamma$-ray lines. The first is $\chi\bar\chi \rightarrow \gamma\gamma$, where
the photon energy is $\sim m_{\chi}$ and the second is $\chi\bar\chi \rightarrow Z^0\gamma$,
where the photon energy satisfies $\varepsilon_{\gamma} = (m_{\chi} - m_Z^2/4m_{\chi})$. The
latter process is only important if the neutralino mass is higher than $\sim$ 45 GeV.

The prediction of $\gamma$-ray fluxes require two independent inputs: that coming from
particle physics for issues such as the interaction cross section and the number of photons
per annihilation, and the input from astrophysics for problems such as the spatial
distribution of dark matter in potential sources.

Here we present some results and predictions based on our previous work [29]. For the sake
of completeness, we summarize here the main assumptions of these calculations (the
reader is referred to [29] for more details):
a) neutralinos are initially supposed to be in thermal equilibrium
with the cosmic plasma; b) they are non-relativistic at decoupling and their relative 
abundance at the freezing point should provide a relic density, corresponding to $\Omega_m \approx$ 0.26;
c) the number of photons per annihilation is estimated from fragmentation functions of
QCD jets of energy $\sim m_{\chi}$. Under these simplified conditions, the neutralino mass is
the sole free parameter. For masses in the range 10$ \leq m_{\chi} \leq$ 2000 GeV, the
decoupling temperature varies within the interval 0.4 - 70 GeV and the thermally averaged
annihilation reaction rate $<\sigma_{\chi\bar\chi}v>$ varies very little,
namely, (7.7 - 9.5)$\times 10^{-27}\, cm^3s^{-1}$, if the ``s-wave" term only is considered.

The galactic center is a privileged potential source of $\gamma$-rays due to its proximity
and high column density. However, the $\gamma$-ray emission from this direction is highly
contaminated by the local background, mostly produced by cosmic ray interactions with
the interstellar environment. In the energy range 0.1 - 1.0 GeV, cosmic ray electrons
produce high-energy photons either by inverse Compton scattering or bremsstrahlung, while
the proton component produces $\gamma$-photons via the decay of neutral pions generated in collisions
with interstellar matter.

EGRET data analyses suggest that, at high galactic latitudes, there is
a residual intensity of $10^{-7} - 10^{-6}$ ph cm$^{-2}s^{-1}sr^{-1}$ above 1 GeV, even
after correction for the expected background of cosmic rays and the diffuse extragalactic
emission [30]. The expected $\gamma$-ray intensity for energies above 1 GeV at $\mid b\mid$ = 90$^o$ 
as a function of the neutralino mass is shown in fig. 1.   

%%%%%%%%%%%%%%%%%%%%%%%%%%%%%%%%%%%%%%%%%%%%%%%%%%%%%%%%%%%%%%%%%%%%%%%%%%%%%%%%%%%%%%
\EFigure{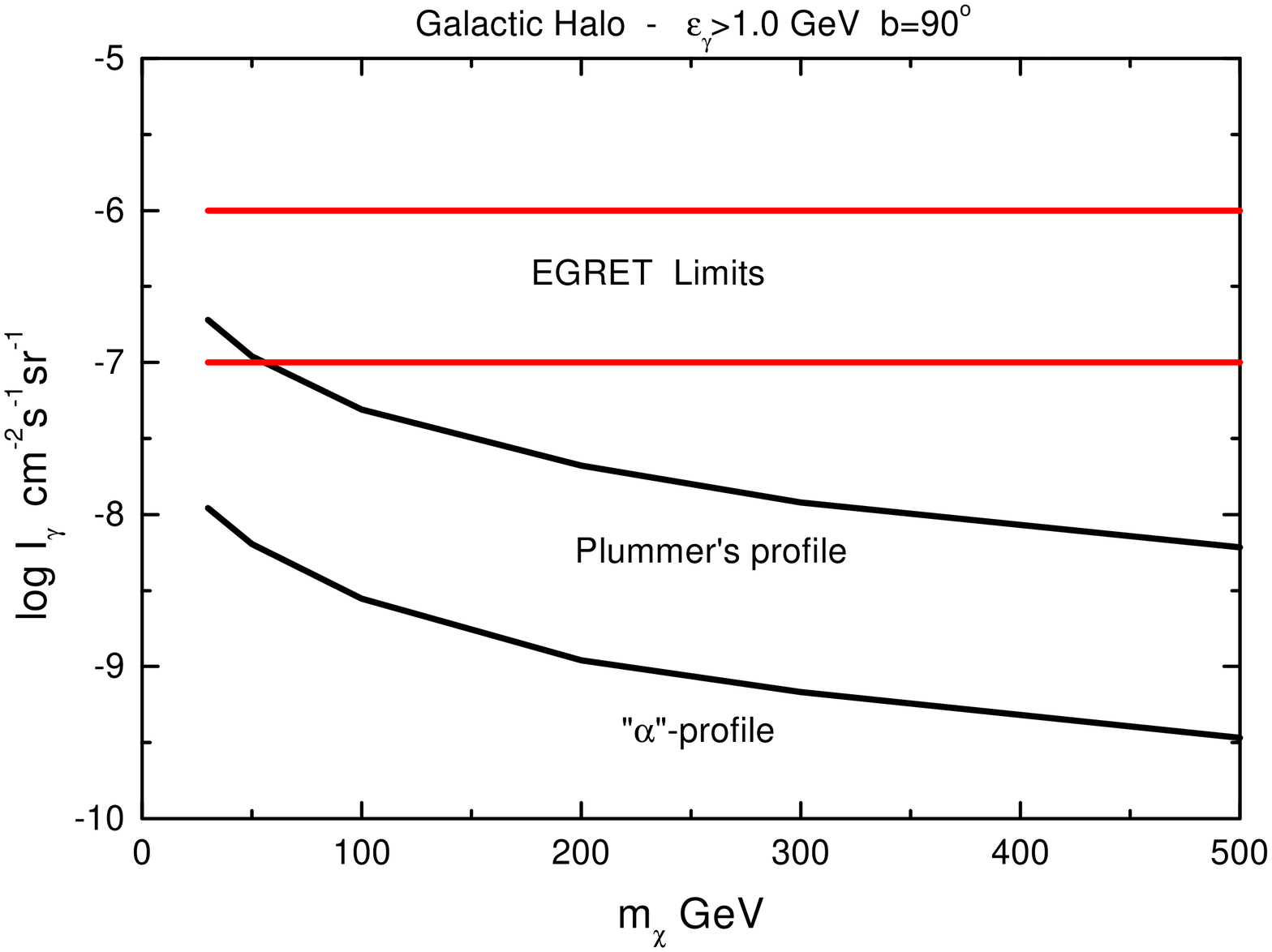}{Predicted $\gamma$-ray intensity above 1 GeV ($\mid b\mid = 90^o$) as
a function of the neutralino mass and for two density profiles. EGRET limits are also
given.}
%%%%%%%%%%%%%%%%%%%%%%%%%%%%%%%%%%%%%%%%%%%%%%%%%%%%%%%%%%%%%%%%%%%%%%%%%%%%%%%%%%%%%%%%

The expected $\gamma$-ray intensity is calculated from the equation

\begin{equation}
I_{\gamma}(r_p) = \frac{<\sigma_{\chi\bar\chi}v>}{4\pi m_{\chi}^2}Q_{\gamma}{\cal I}(r_p)
\end{equation}

In the above equation, $Q_{\gamma}$ is the number of photons produced per
annihilation with energies higher than a given value [29] and ${\cal I}(r_p)$ is the
{\it reduced intensity} at a given projected distance $r_p$ from the center, defined as

\begin{equation}
{\cal I}(r_p) = \int\rho_{\chi}^2(\sqrt{s^2 + r_p^2})ds
\end{equation}
and the integral should be performed along the line of sight.

Two density profiles were considered in the calculations. The first, the recently proposed
 ``$\alpha$"-profile [31], which provides a {\it finite} central 
density and is able
to fit adequately the inner structure of dark halos resulting from numerical simulations, namely,
\begin{equation}
\rho(r) = \rho_*exp\lbrace -\frac{2}{\alpha}\lbrack (\frac{r}{r_*})^{\alpha} - 1\rbrack\rbrace
\end{equation}

The second is a Plummer profile, intended to represent better the baryon-to-dark matter ratio
resulting from analyses of rotation curves of bright galaxies, e.g.,
\begin{equation}
\rho(r) = \frac{\rho_0}{\lbrack 1 + (1/3)(r/r_0)^2\rbrack^{5/2}}
\end{equation}

For the Galaxy, the parameters defining the aforementioned density profiles are:
$\rho_* = 0.0061\, M_{\odot}pc^{-3}$, $r_*$ = 11.6 kpc, $\alpha = 0.17$
(``$\alpha$"-profile)
and $\rho_0 = 0.038\, M_{\odot}pc^{-3}$, $r_0$ = 12.2 kpc (Plummer profile) [29].

Inspection of fig. 1 shows that the EGRET residual emission can be explained if the neutralino
mass is less than 50 GeV, a value marginally consistent with the LEP lower limit. Contrary
to what is generally obtained when the galactic center direction is considered, at high
latitudes the Plummer profile predicts an intensity {\it higher} than that derived
from a ``cuspy" density profile. This is easily understood since the latter profile gives
a larger mass concentration near the center while the former has a shallower mass
distribution. At high galactic latitudes, the predicted intensities from 
the ``$\alpha$"-profile are always below the
EGRET residual values. In this case, an important enhancement by substructures in the halo
is required. However, the expected enhancement factor, according to numerical simulations
performed by [29], is rather
small, not exceeding a factor of 2. Presently, a firm conclusion cannot be made since
the EGRET residuals are in the sensibility limit of the instrument. The situation is
expected to improve greatly with the forthcoming Gamma-ray Large-Area Space Telescope (GLAST).

The study of $\gamma$-ray emission with GLAST has some advantages over atmospheric Cherenkov
telescopes: i) lower energy threshold, allowing to probe neutralino masses above 10 GeV; ii)
the background is mainly due to the diffuse extragalactic emission, and iii) the spatial
resolution varies with the threshold energy, permitting us to probe also the density profile.

Here we consider two energy thresholds: 0.1 and 1.0 GeV. Then, we compare the predicted
$\gamma$-ray intensities for M31 and M87 as a function of the neutralino mass with the
detectability limit of GLAST. Parameters defining the halo properties of these galaxies
are the same as [29]. Both objects have probably a massive black hole in their centers [32],
which boost significantly the $\gamma$-emission by producing a central density spike
within their sphere of influence. This effect was included when $\gamma$-ray fluxes were
computed. 

The result of combining the information on both aforementioned energy
thresholds is summarized in table 1. Columns two and three indicate if the galaxy is
detected or not at the corresponding energy threshold and consequences for the expected
density profile (column four): ``cored" or ``cuspy". Finally, column five gives the
neutralino mass range expected from a positive or negative detection by GLAST.
%%%%%%%%%%%%%%%%%%%%%%%%%%%%%%%%%%%%%%%%%%%%%%%%%%%%%%%%%%%%%%%%%%%%%%%%%%%%%%%%%%%%%%%
{\small
\begin{table}
\caption{{\small Neutralino masses from positive or negative detection of
M31 and M87 by GLAST}}
\begin{flushleft}
\begin{tabular}{lccccc}
\noalign{\smallskip}
\hline
\noalign{\smallskip}
Object&$E_{\gamma} >$ 0.1 GeV&$E_{\gamma} >$ 1.0 GeV& profile&$m_{\chi}$ (GeV) \\
\noalign{\smallskip}
\hline
\noalign{\smallskip}
M31&no&no&cored&$>$ 20\\
M31&yes&yes&cuspy&$<$ 300\\
M31&yes&no&cored&$<$ 20\\
M31&no&yes&cuspy&300-500\\
M87&no&no&cored&$>$ 100\\
M87&yes&yes&cuspy&$<$ 60\\
M87&yes&no&cored& - \\
M87&no&yes&cuspy& 60-100\\
\noalign{\smallskip}
\hline
\end{tabular}
\end{flushleft}
\end{table}
}
%%%%%%%%%%%%%%%%%%%%%%%%%%%%%%%%%%%%%%%%%%%%%%%%%%%%%%%%%%%%%%%%%%%%%%%%%%%%%%%%%%%%%%%%%%%%%

\section{Warm dark matter...?}

Presently, the cold dark matter paradigm explains successfully the large-scale structure 
in the galaxy distribution on scales of 0.02 $< k <$ 0.15h Mpc$^{-1}$ [33,34]. The dark
matter power spectrum on these scales derived from large redshif surveys as, for instance,
the Anglo-Australian 2-degree Field Galaxy Redshift Survey (2dFGRS), is also consistent with
the Lyman-$\alpha$ forest data in the redshift range 2$< z < $4 [35,36,37].

In spite of these impressive successes, there are still discrepancies 
between simulations and observations at scales $\leq$ 1 Mpc. The first problem concerns the 
sharp central density cusp of dark matter halos predicted by simulations and 
not seen in the rotation curves of bright spiral galaxies [38,39].
The second difficulty is related to the large number of sub-halos present in simulations but
not observed [40,41], as in the case of our Galaxy or M31. Besides these difficulties,
deep surveys ($z \geq 1-2$) as the Las Campanas Infrared Survey, HST Deep Field
North and Gemini Deep Deep Survey (GDDS) are revealing an excess of massive galaxies with
respect to predictions of the hierarchical scenario [42].   

These problems could be alleviated if dark matter particles had a free streaming (or Landau
damping) length-scale higher than usually supposed. In this case, the smearing out of the small scale
structure could bring simulations in better agreement with observations, solving some
of the difficulties mentioned above [43]. Particles decoupling relativistically but having
became non-relativistic {\it before} the matter-radiation equality, constitutes the so-called
``warm" dark matter. These particles have a
velocity dispersion higher than neutralinos when structures began to be formed, thus
filtering density perturbations at a higher cut-off.   

Sterile neutrinos are a possible ``warm" dark matter particle candidate [44]. These particles
are Standard Model singlet fermions, which couple to the conventional (``active") neutrinos
($\nu_e\nu_{\mu}\nu_{\tau}$) solely via effective mass terms and are neutral under
all Standard Model gauge forces. Very massive sterile neutrinos arise naturally in the
so-called ``see-saw" models in Grand Unified Theories (GUTs) [45].

Besides their interest for cosmology, sterile neutrinos have been proposed to solve
the apparent discrepancies between the neutrino mass-squared differences ($\delta m_{\nu}^2$)
resulting from several experiments and now explained simply in terms of oscillations among
the three active neutrinos [46]. The conversion of active into sterile neutrinos has also
been invoked to solve ``anemic" r-process nucleosynthesis in supernova ejecta, resulting from
neutrino-driven shocks. Such a conversion reduces the electron
number per baryon Y$_e$, favoring the nucleosynthesis of heavy (and neutron rich) elements [47].  
 
In order to be an acceptable dark matter candidate, sterile neutrinos must satisfy some requirements:
i) they must be able to produce the observed relic density; ii) their abundance
should not alter the results of big-bang nucleosynthesis; iii) they should obey the
constraints imposed by the core collapse of SN1987A and, finally, iv) have a lifetime 
longer than H$_0^{-1}$. Some of these requirements have recently been reviewed in [48].

Taking into account different constraints, it is possible to define a region relevant
for cosmology in the plane ``mass-mixing angle" (Fig. 2). 
Based on calculations performed by [49], curves 1 and 2 define respectively
the region where these parameters satisfy the condition $\Omega_mh_0^2$ = 0.11 and
the region disfavored by considerations on the supernova core collapse. The
resulting density of sterile neutrinos (curve 1) was estimated from non-equilibrium
processes in the early universe, taking into account incoherent resonant
and non-resonant scattering and an initial lepton asymmetry L$_{\nu}$ = 0.01 [49]. 

%%%%%%%%%%%%%%%%%%%%%%%%%%%%%%%%%%%%%%%%%%%%%%%%%%%%%%%%%%%%%%%%%%%%%%%%%%%%%%%%
\EFigure{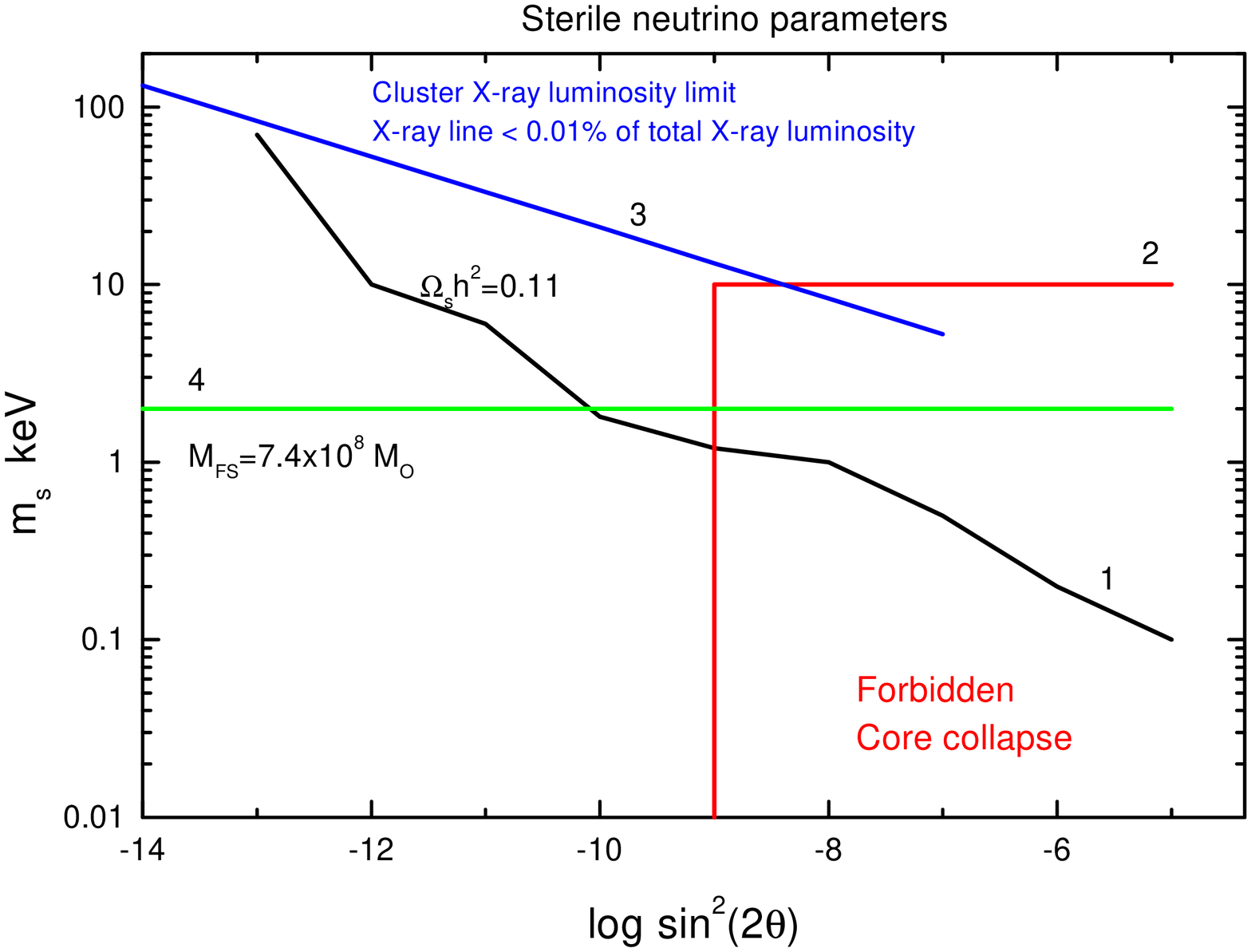}{Allowed masses and mixing angles for sterile neutrinos}
%%%%%%%%%%%%%%%%%%%%%%%%%%%%%%%%%%%%%%%%%%%%%%%%%%%%%%%%%%%%%%%%%%%%%%%%%%%%%%%%

Sterile neutrinos, if they do not feel Standard Model gauge interactions only, are labeled
``weakly sterile" whereas if they do not feel {\it any} gauge interaction (including those beyond
the Standard Model), they are dubbed ``fully sterile". In the former case, they can decay into
lighter ``active" neutrinos or radiatively, with a decay branch ratio
$\Gamma(\nu_s \rightarrow \nu_a\gamma)/\Gamma(\nu_s \rightarrow 3\nu_a) = 27\alpha/8\pi$, 
the photon energy satisfying $\epsilon_{\gamma} \sim \frac{1}{2}m_{\nu_s}$. This
X-ray line emission can be considered as a possible signature of keV-sterile neutrinos.
Curve 3 in fig. 2 results from the assumption that in a typical cluster of galaxies of mass of 
about 10$^{14}\, M_{\odot}$, the X-ray line flux produced by the decay of sterile neutrinos
is of the order of 10$^{-4}$ of the continuum emission due to the hot gas. Finally, curve 4
indicates the ``free-streaming" mass ($M_{FS} = 7.4\times 10^8\, M_{\odot}$)  for a 2 keV sterile 
neutrino. Lower sterile neutrino masses will excessively increase  $M_{FS}$, destroying
the agreement between theory and observations at large scales. According to fig. 2, masses
up to 10-20 keV are still allowed, but then the $M_{FS}$ scale is so low that practically
no differences from cold particle dynamics exist.

It is worth mentioning that sterile neutrinos in the mass range 1-20 keV and with
comparable mixing angles could also be able to explain the origin of the natal kick of
pulsars [50], which could be an additional point in favor of their existence.  

\section{... or Light dark matter ?}

Recent observations with the spectrometer SPI on board of the space observatory INTEGRAL have not
only confirmed past detections of the 511 keV line emission from the galactic center, but have also 
revealed the extended nature of the emission [51]. This emission is the indisputable signature
of electron-positron pair annihilations. Possible astrophysical sources of positrons as neutron
stars, black holes, novae, type Ia supernovae fall short of explaining the measured line
intensity ($9.9\times 10^{-4}\, cm^{-2}s^{-1}$). Positrons can be generated
in the neutralino annihilation process through different channels (see eq. 1). In particular,
charged and neutral pions, produced roughly at the same number, will decay ultimately 
into $e^+, e^-$ and photons respectively. Thus, if
positrons in the galactic center are originated from $\chi\bar\chi$ annihilations, a $\gamma$-ray flux 
($\epsilon_{\gamma} > $ 60 MeV) {\it higher} than EGRET upper limits would have been observed, which
is not the case.   
As a consequence, several alternative scenarios involving either annihilation or decaying
dark matter particles have been proposed to explain the 511 keV emission. 

Decaying axinos, with masses in the range 1-300 MeV, in an R-parity violating model of supersymmetry
could be possible candidates, producing positrons through the channels, $\tilde a \rightarrow
\nu_{\tau}e^+e^-$ or $\tilde a \rightarrow \nu_{\mu}e^+e^-$ [52]. In this scenario, axinos
constitute the major dark matter component and might be present in the galactic halo with a cusped
density profile such as $dlg\rho/dlg r \sim$ 1.2 .

Weakly sterile neutrinos were also proposed as a source for positrons by means of the
decay channel $\nu_s \rightarrow \nu_ae^+e^-$ [53]. In this case, masses of the
sterile neutrino are in the range 1-50 MeV. A negative aspect of this scenario is that the
required mixing angles consistent with the desired mass interval lead to cosmic densities
of $\Omega_{\nu_s}h^2 \sim 10^{-6}$. Thus, it is not possible to explain simultaneously
the 511 keV emission {\it and} the cosmic dark matter density.

A rather different approach was followed in [54]. New 0-spin  MeV relic particles are 
postulated, feeling a force field carried by a new light gauge boson {\it U}. Positrons
would be generated almost at rest via annihilation, e.g., $X\bar X \rightarrow e^+e^-$.
If the annihilation cross section is velocity dependent and if the density profile of
the galactic halo have a central slope $dlg \rho/dlg r \sim$ 0.6, then it is possible to
explain the observed spatial profile of the 511 keV line emission and to obtain
a concentration consistent 
with the relic dark matter density [55,56].   

MeV thermal relic particles are expected to be coupled to the cosmic plasma at the epoch
of big-bang nucleosynthesis and thus, they might contribute to the energy density and
expansion rate. If, during nucleosynthesis, X-particles are mainly coupled to neutrinos,
then their masses should be {\it higher} than 10 MeV in order not to alter the predicted
abundances of $^2$H, $^4$He and $^7$Li. If the coupling is essentially electromagnetic,
X-particles in the mass range 4-10 MeV can even improve slightly the agreement between predicted
and observed abundances [57]. 

Mass limits for dark matter particles can also be obtained from the power spectrum of primordial 
fluctuations at small scales ($< 1h^{-1}$\,Mpc) derived from Lyman-$\alpha$ absorbers present
in the spectra of quasars [58]. However, these limits depend on the adopted reionization epoch.
Data on high-redshift quasars suggest that reionization occurred at $z\sim 6$, implying
dark matter particles with masses around 1-5 keV [59], whereas WMAP data favor an earlier
epoch ($z \sim 20$), implying particle masses $\geq$ 1 MeV [1,58]

\section{Conclusions}

New EROS data combined with previous microlensing observations were analyzed in [60], leading
to an upper limit of about 10\% for MACHOs in the mass range $10^{-6} - 0.3\, M_{\odot}$, able to
contribute
to the total mass of the galactic halo. This result suggests that most of the halo
dark matter must probably be under the form of elementary particles.

Supersymmetric particles are privileged candidates. Several experiments are underway for 
the direct detection of WIMP particles. Up today, only the DAMA collaboration claims for a
positive detection of a modulated signal compatible with a particle mass of 52$\pm$10 GeV
and a WIMP-nucleon cross section of about 7$\times 10^{-6}$ picobarn. No satisfactory
explanations have been found to explain the nature of such a signal face to negative
results of other experiments [61]. 

Searches for energetic neutrinos resulting from neutralino annihilations in the center
of the Earth or in the center of the Sun impose only upper limits on the fluxes. The
comparison of these limits with theoretical expectations is still quite doubtful, since
uncertainties present in the calculations of the capture rate of WIMPs by the Earth were
not completely removed yet [21]. Antiprotons in the energy range 0.18-1.4 GeV detected
by BESS collaboration, if produced in the follow up hadronization of neutralino annihilations,
imply masses m$_{\chi} <$ 100 GeV.

$\gamma$-rays resulting from $\pi^o$ decay, formed in the hadro-nization process,
are a promising possibility of indirect detection of dark matter. Searches for very high
energy photons via atmospheric Cherenkov telescopes such as VERITAS, CELESTE, MAGIC, have
not revealed any positive signal yet. The quite uncertain EGRET residual emission seen
at high galactic latitudes above 1 GeV could be explained by neutralino annihilations
if: i) the dark matter profile is ``cored" and ii) the neutralino mass is $\leq$ 50 GeV.
Notice that this mass limit is compatible with those derived from antiproton data analysis and 
from LEP data. Detection or upper limits on $\gamma$-ray fluxes from potential sources
as M31 or M87, at different energy thresholds by the forthcoming GLAST, will improve
considerably limits on the neutralino mass and will shed some light on the
their spatial distribution inside galactic halos.

Difficulties with {\it cold} dark matter at small scales lead to alternative
scenarios as {\it warm} particles, whose a possible candidate is a sterile neutrino in the 
keV mass range. These particles
 provide an adequate free streaming mass able to solve {\it some} small scale
problems and are not in conflict with X-ray data from galaxy clusters. Moreover, they
provide also a natural mechanism to explain the natal kick of pulsars. However, structures
in a warm dark matter universe appear lately in comparison with a cold dark matter model, being
a difficulty to form early sources responsible for the reionization of the universe evidenced
by WMAP. 

Finally, the extended nature of the 511 keV line emission from the galactic center revealed
by INTEGRAL observations, raised the possibility of the existence of MeV dark matter particles,
feeling a new gauge force field. These particles will not affect the primordial nucleosynthesis
but, from a dynamical point of view, they will have the same difficulties at small scales as 
heavy particles.

%%%%%%%%%%%%%%%%%%%%%%%%%%%%%%%%%%%%%%%%%%%%%%%%%%%%%%%%%%%%%
%%%%%%%%%%%%%%%%%%%%%%%%%%%%%%%%%%%%%%%%%%%%%%%%%%%%%%%%%%%%%%%%%%%%%%%%%%%%%%%%%%%
\Acknow
{S.P. acknowledges the University of Nice-Sophia Antipolis for the financial support}
%%%%%%%%%%%%%%%%%%%%%%%%%%%%%%%%%%%%%%%%%%%%%%%%%%%%%%%%%%%%%%%%%%%%%%%%%%%%%%%%%%%%%

\end{document}